\documentclass[aps,pre,twocolumn,10pt,amsmath,amssymb,superscriptaddress,longbibliography,floatfix]{revtex4-1}

\usepackage{amsmath}
\usepackage{amssymb}
\usepackage{empheq}
\usepackage[parfill]{parskip}
\usepackage{color}
\usepackage{array}
\usepackage{booktabs}
\usepackage{amsfonts}
\usepackage{graphicx}
\usepackage{mathtools}
\usepackage{natbib}
\usepackage{xcolor}
\usepackage{placeins}
\usepackage{hyperref}

\begin{document}

\setlength{\parindent}{0.5cm}

\title{Coupling-split clusters in a swarmalator model with uniform coupling disorder}

\author{K. P. O'Keeffe}
 \email{kevin.p.okeeffe@gmail.com}
 \affiliation{Starling Research Institute, Seattle, WA 98112, USA}

\begin{abstract}
We study the one-dimensional swarmalator model in which the phase coupling $K_i'$ is drawn from a uniform distribution. Our main result is a static coupling-split cluster, in which the population partitions across the threshold $K'=0$ that separates positively coupled ($K_i'>0$) from negatively coupled ($K_i'<0$) swarmalators, with smaller order parameter $s=\mu/\gamma$ set by the positive-coupling excess. The familiar async, phase-wave, and sync states persist, but each stability boundary feels a different part of the distribution: async the mean same-coordinate response, sync the most negatively coupled particle, and the phase wave the full density through a logarithmic characteristic equation. At a cusp where its Hopf and real-eigenvalue branches meet, the phase-wave dispersion has a double zero -- the spectral signature of a Bogdanov--Takens point -- and simulations nearby show a small-amplitude breathing limit cycle. For supports containing strongly negatively coupled particles the order parameters instead oscillate persistently.
\end{abstract}

\maketitle

\section{Introduction}
Swarmalators are mobile oscillators whose spatial aggregation and phase synchronization are coupled \cite{o2022collective,yoon2022sync}. They provide minimal models for systems in which self-assembly and synchronization interact, including vinegar eels \cite{quillen2021metachronal}, tree frogs \cite{aihara2014spatio}, magnetic domain walls \cite{hrabec2018velocity}, colloidal micromotors \cite{yan2012linking,liu2021activity,zhang2020reconfigurable,bechinger2025tunable,snezhko2011magnetic}, embryonic cells \cite{tsiairis2016self}, active spheres \cite{riedl2023synchronization}, hydrodynamically entrained and dense biological suspensions \cite{riedel2005self,creppy2016symmetry}, robotic swarms \cite{barcis2019robots,barcis2020sandsbots,xu2026navigation}, and human interactions on dance floors \cite{toiviainen2025modeling}.

The one-dimensional model is especially useful because its symmetry allows exact or semi-exact calculations while preserving the basic feedback between motion and phase order \cite{o2022collective,yoon2022sync,global_sync,o2025stability}. Within and beyond this solvable setting, recent work has used swarmalator models as a test bed for coupling disorder \cite{o2022swarmalators,hao2023attractive}, phase frustration and phase lag \cite{lizarraga2023synchronization,senthamizhan2025frustration,sharma2025phaseLag,sharma2026forcedPhaseLag,sharma2026resonantPhaseLag}, directed or nonreciprocal coupling \cite{yu2025directed}, random pinning \cite{sar2023pinning,sar2024solvable,sar2023swarmalators}, external forcing \cite{anwar2024forced,anwar2025forced}, thermal noise \cite{hong2023swarmalators}, finite-range and topological interactions \cite{sar2025effects,gou2026topological}, higher-order interactions \cite{anwar2024collective,anwar2025twoDimHigherOrder}, inertia \cite{okeeffe2026inertia}, time delay \cite{okeeffe2026delay}, and other swarmalator extensions \cite{ghosh2025dynamics,ghosh2026emergent,Senthamizhan2026FrequencyWeighted,louodop2025topological,Acharya2025WinfreeSwarmalator,o2024solvable,schilcher2025multicircular,lee2025circular}. 

Coupling heterogeneity is a natural next ingredient to study. In ordinary oscillator theory, allowing positive and negative phase couplings gives the conformist--contrarian problem, where even mean-field Kuramoto variants can produce antiphase locking, traveling waves, and clustered dynamics \cite{hong2011kuramoto,manoranjani2024asymmetric}. In one-dimensional swarmalators the same issue is richer because coupling disorder also acts through the swarmalator coupling channel. Previous studies of swarmalators with mixed-sign coupling assumed, for simplicity, binary couplings \cite{o2022swarmalators,hao2023attractive}. This paper studies a tractable continuous extension of that setting. We let the phase coupling $K_i'$ vary across particles and take the uniform distribution,
\begin{align}
h(K')=
\begin{cases}
\dfrac{1}{2\gamma}, & \mu-\gamma\leq K'\leq\mu+\gamma,\\
0, & \text{otherwise},
\end{cases}
\label{uniform-density}
\end{align}
which is simple enough to allow exact results while capturing the key feature of a continuous spread of coupling strengths.

The main new result is a static state we call the coupling-split cluster, which appears when the support straddles the threshold $K'=0$ -- the sign change that separates positively coupled ($K_i'>0$) from negatively coupled ($K_i'<0$) particles. We also observe unsteady dynamics when the support extends below $K'=-1$ (i.e., contains strongly contrarian particles), in which the order parameters oscillate persistently. We characterize the static states analytically and describe the unsteady regimes numerically.

\section{Model}
We consider $N$ swarmalators on a ring, each with position $x_i\in S^1$ and phase $\theta_i\in S^1$. The equations of motion are
\begin{align}
\dot x_i &= \nu_i + \frac{1}{N}\sum_j J'\sin(x_j-x_i)\cos(\theta_j-\theta_i),\\
\dot\theta_i &= \omega_i + \frac{1}{N}\sum_j K_i'\sin(\theta_j-\theta_i)\cos(x_j-x_i),
\end{align}
where $J'=1$ is a fixed homogeneous spatial coupling and $K_i'$ is a heterogeneous phase coupling drawn from \eqref{uniform-density}, with $K_i'\in[\mu-\gamma,\mu+\gamma]$. We set $\nu_i=\omega_i=0$. The limit $\gamma=0$ recovers the identical-coupling 1D swarmalator model with $J'=1,\ K'=\mu$~\cite{o2022collective,yoon2022sync}.

In sum and difference coordinates
\begin{align}
\xi_i = x_i+\theta_i,\qquad \eta_i = x_i-\theta_i,
\end{align}
the trigonometric identity $\sin a\cos b = \tfrac12[\sin(a+b)+\sin(a-b)]$ recasts the dynamics in terms of two `response coefficients'
\begin{align}
a_i = \frac{1+K_i'}{2},\qquad b_i = \frac{1-K_i'}{2},\qquad a_i+b_i = 1,
\label{eq:ab-def}
\end{align}
where $a_i$ is the same-coordinate response and $b_i$ the cross-channel response. With the order parameters
\begin{align}
r e^{i\phi} = \left\langle e^{i\xi}\right\rangle,\qquad
s e^{i\psi} = \left\langle e^{i\eta}\right\rangle,
\end{align}
($\langle\cdot\rangle = N^{-1}\sum_j$), the equations of motion become
\begin{align}
\dot{\xi}_i &= a_i\, r\sin(\phi-\xi_i) + b_i\, s\sin(\psi-\eta_i), \label{eq:uniform-xi}\\
\dot{\eta}_i &= b_i\, r\sin(\phi-\xi_i) + a_i\, s\sin(\psi-\eta_i). \label{eq:uniform-eta}
\end{align}
A key structural feature of this model is that not only the same-coordinate response $a_i$ but also the cross-channel response $b_i$ is heterogeneous; the constraint $a_i+b_i=1$ encodes the homogeneous spatial coupling $J'=1$. This distinguishes the model from conformist--contrarian Kuramoto, where heterogeneity enters only the same-coordinate self-coupling \cite{hong2011kuramoto}.

In the continuum limit the state is a density $\rho(\xi,\eta,K',t)$ and the order parameters are
\begin{align}
r e^{i\phi}
&=\iiint e^{i\xi}\rho(\xi,\eta,K',t)h(K')\,d\xi\, d\eta\, dK',\\
s e^{i\psi}
&=\iiint e^{i\eta}\rho(\xi,\eta,K',t)h(K')\,d\xi\, d\eta\, dK'.
\end{align}
We use these to classify the system's macrostates.

\section{Numerics}
\label{sec:numerics-states}

\begin{figure*}[!t]
\centering
\includegraphics[width=0.96\textwidth]{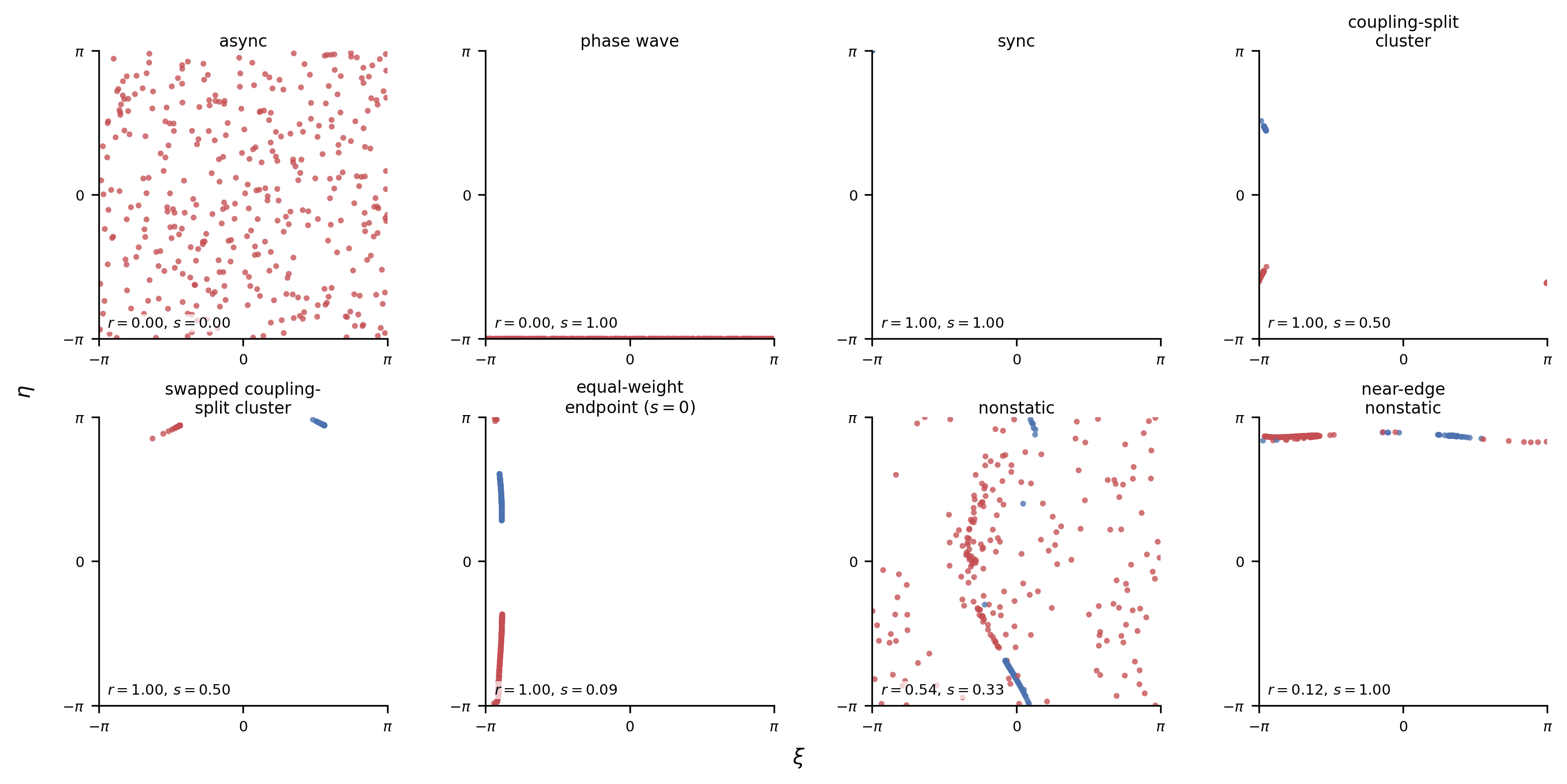}
\caption{State gallery for the uniform phase-coupling model in the $(\xi,\eta)$ plane. Points are colored by the sign of $K_i'$: blue for $K_i'>0$ (conformist) and red for $K_i'<0$ (contrarian). The first row shows async ($\mu=-1.5$, $\gamma=0.5$), phase wave ($\mu=-0.3$, $\gamma=0.2$), sync ($\mu=0.7$, $\gamma=0.3$), and a coupling-split cluster ($\mu=0.2$, $\gamma=0.5$). The second row shows the coordinate-swapped coupling-split cluster (same parameters), the equal-weight endpoint ($s=0$ at $\mu=0$, $\gamma=0.6$), the strongly-contrarian unsteady state ($\mu=-0.3$, $\gamma=1.5$), and a snapshot of the small-amplitude periodic orbit near the BT cusp ($\mu=0.05$, $\gamma=0.80$) discussed in \S\ref{ssec:phase-wave}.}
\label{fig:uniform-gallery}
\end{figure*}

We simulated Eqs.~\eqref{eq:uniform-xi}--\eqref{eq:uniform-eta} using deterministic quantiles of the uniform distribution for $K_i'$. Figure~\ref{fig:uniform-gallery} shows representative particle-level states in the $(\xi,\eta)$ plane, colored by the sign of $K_i'$.

We found the following states:

(i) \emph{Async.} The async state fills the $(\xi,\eta)$ torus and has $r=s=0$ up to finite-size fluctuations. Coupling signs are spatially mixed because the state is incoherent in both coordinates. In the analysis below, async's stability is controlled only by the mean same-coordinate response $\langle a\rangle=(1+\mu)/2$.

(ii) \emph{Sync.} The sync state has both split coordinates locked, so $r=s=1$. In Fig.~\ref{fig:uniform-gallery} we use rotational freedom to place the synchronized point at $(\xi,\eta)=(0,0)$. Its stability is controlled by the weakest coupling in the population, hence by the lower edge $\mu-\gamma$.

(iii) \emph{Phase wave.} The phase wave has one split coordinate locked and the other uniformly distributed. In the displayed orientation, $\eta=0$ and $\xi$ winds around the ring, giving $s=1$ and $r=0$. Unlike async and sync, the phase-wave stability boundary depends on the full distribution of $K_i'$.

(iv) \emph{Coupling-split clusters.} When the support of $h(K')$ straddles $K'=0$ and no particle is too strongly contrarian ($\mu-\gamma>-1$), the population splits into two clumps selected by the sign of the phase coupling: conformist particles ($K_i'>0$) occupy one location and contrarian particles ($K_i'<0$) occupy the opposite location. This gives $r=1$ (all particles aligned in $\xi$) and $0<s<1$ (the $\eta$ coordinate split between two antipodal clumps). The smaller order parameter $s$ equals the conformist excess. The coordinate-swapped orientation has $0<r<1$ and $s=1$.

(v) \emph{Nonstatic and multistable states.} Strongly-contrarian supports (those containing particles with $K_i'<-1$) produce nonstatic trajectories in which neither split coordinate settles to a simple static branch. Near the edge of the phase-wave and coupling-split regions, simulations also show multistability: prepared initial data can remain close to a partially ordered branch while nearby random data move to an unsteady state.

(vi) \emph{Small-amplitude periodic orbit near the phase-wave cusp.} Near the endpoint of the phase-wave Hopf branch at $(\mu_{\rm end},\gamma_{\rm end})\approx(0.219,0.827)$, simulations from random initial conditions reveal a small-amplitude periodic orbit distinct from both the coupling-split cluster and the phase wave. In the displayed orientation it has one split coordinate effectively locked ($s\approx 1$) and the other split into two clumps separated by $\pi$, with the smaller order parameter $r$ pulsing periodically -- a \emph{breathing coupling-split cluster}. The cusp endpoint carries the spectral signature of a Bogdanov--Takens point in the phase-wave dispersion (\S\ref{ssec:phase-wave}), and this orbit is the small-amplitude limit cycle observed there.

The scan used $N=112$\footnote{$N=112=16\times7$ was chosen so that the deterministic quantile grid $K_i'=\mu-\gamma+2\gamma(i+1/2)/N$ divides evenly across a wide range of $\mu/\gamma$ ratios, minimizing rounding in the coupling-split construction. The physics is insensitive to this choice.}, a fourth-order Runge--Kutta method with $dt=0.06$, total time $T=90$, and classified the final $35\%$ of each trajectory. The couplings were the deterministic quantiles $K_i'=\mu-\gamma+2\gamma(i+1/2)/N$, and each parameter pair was tested from random, phase-wave, and synchronized initial conditions. A run was called nonstatic if the late-time mean speed exceeded $0.05$ or if the standard deviation of either order parameter exceeded $0.035$. Static states were separated using $r$, $s$, and the second harmonics $r_2=|\langle e^{2i\xi}\rangle|$, $s_2=|\langle e^{2i\eta}\rangle|$, which distinguish phase waves from two-clump coupling-split states.

The deterministic quantiles are used as a quadrature rule for the continuum distribution, not as an extra structural assumption; iid quenched disorder checks confirming the main states are given in Appendix~\ref{app:iid}.

\begin{figure}[tb]
\centering
\includegraphics[width=\columnwidth]{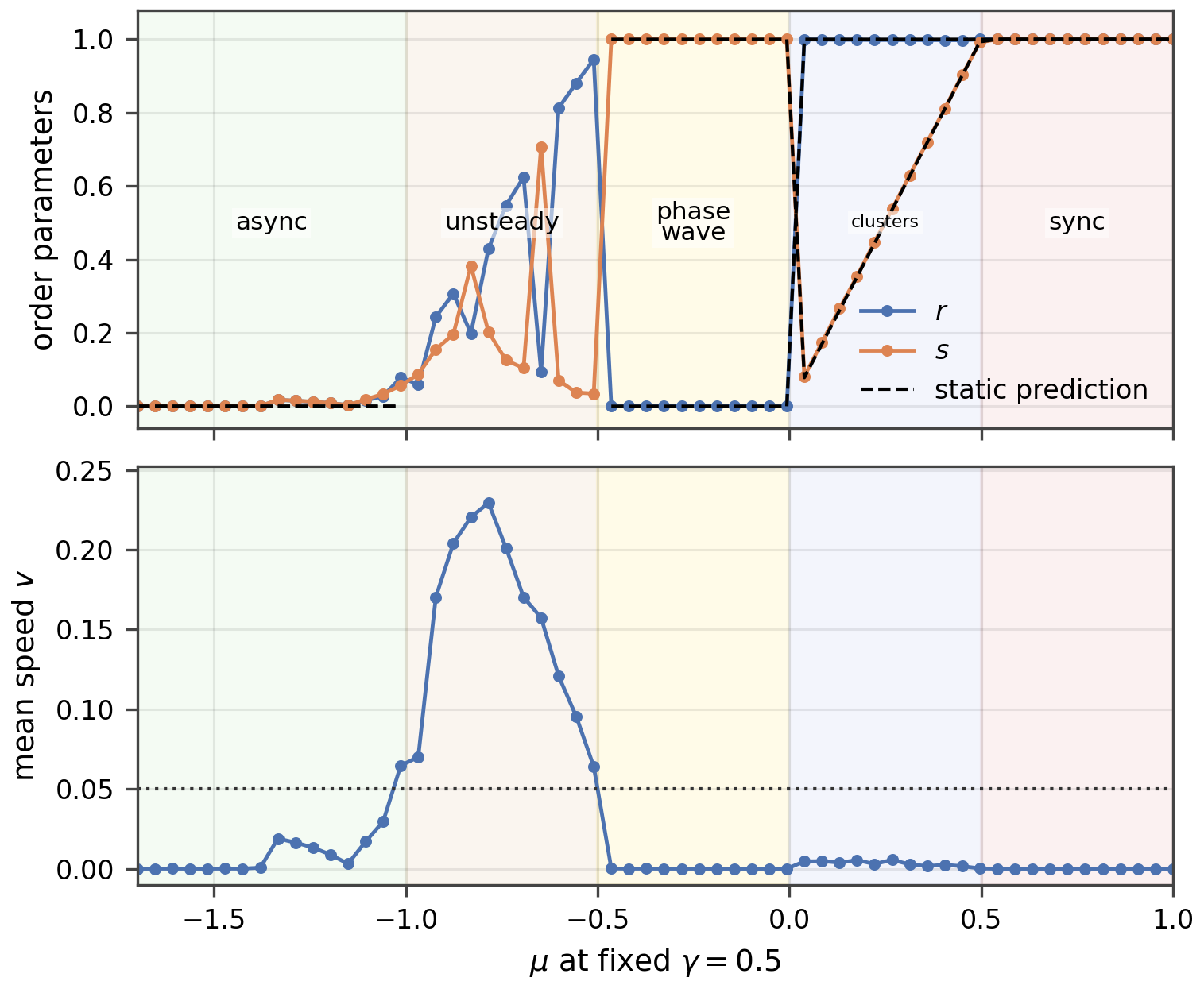}
\caption{Order-parameter section at fixed $\gamma=0.5$. The upper panel shows the order parameters $r=|\langle e^{i\xi}\rangle|$ and $s=|\langle e^{i\eta}\rangle|$; the lower panel shows the late-time mean speed $v$. Because the phase wave is $\eta$-locked ($s=1$, $r=0$) while the coupling-split cluster is $\xi$-dominant ($r=1$, $s=\mu/\gamma$), $r$ and $s$ exchange roles across the transition at $\mu=0$. The dashed curves are the static predictions on the async, phase-wave, coupling-split, and sync branches; the shaded bands indicate the analytic regions.}
\label{fig:uniform-order-mu}
\end{figure}

\begin{figure}[tb]
\centering
\includegraphics[width=\columnwidth]{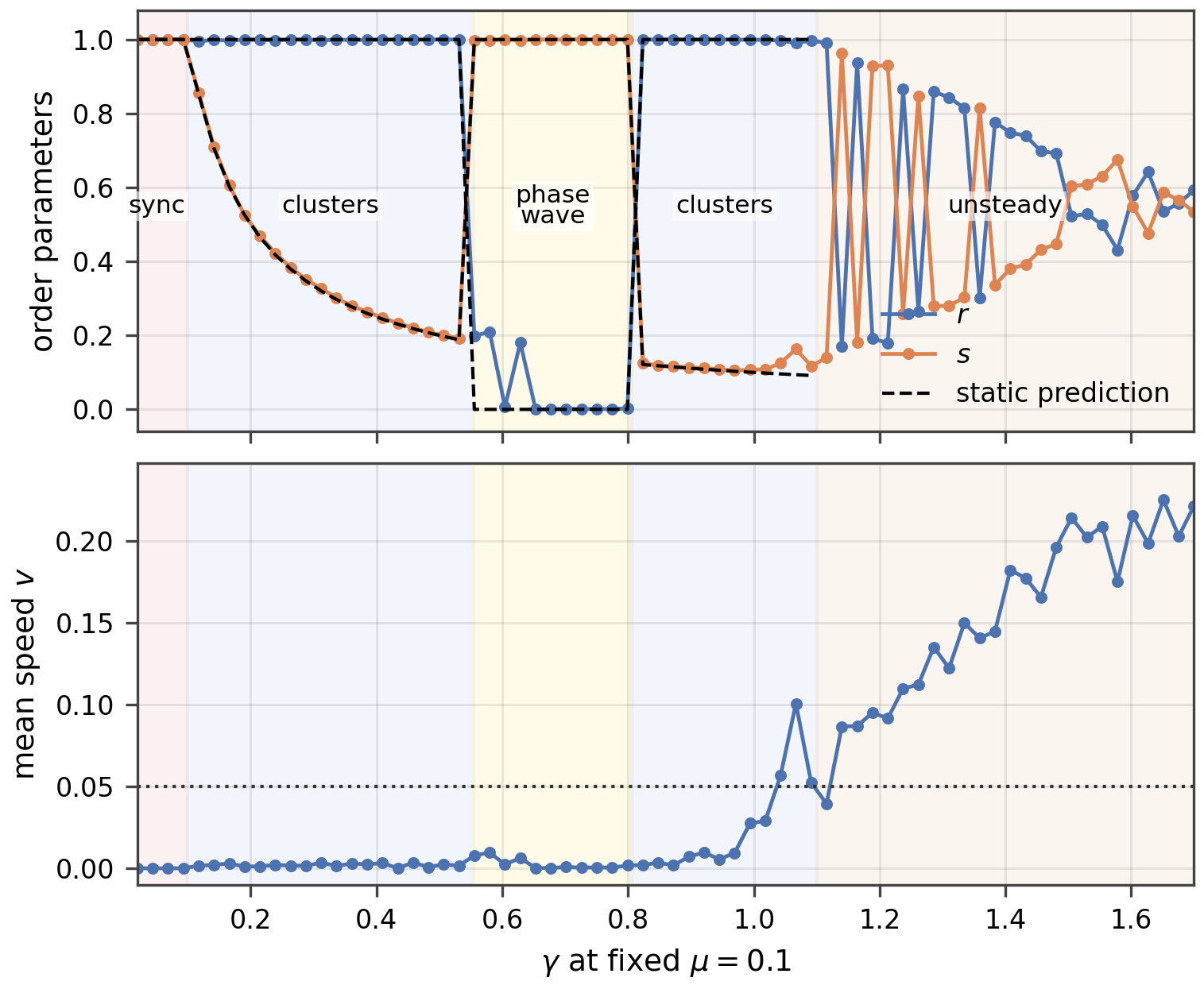}
\caption{Observed-attractor reentrance of the coupling-split cluster. Order-parameter section at fixed $\mu=0.1$ sweeping $\gamma$ from $0$ to $1.7$. The coupling-split construction exists analytically throughout $\mu<\gamma<\mu+1$ ($0.1<\gamma<1.1$); the attractor reached from random initial conditions is the cluster for $\mu<\gamma<0.554$ and again for $0.807<\gamma<\mu+1$, with the phase wave captured instead for $\gamma\in(0.554,0.807)$. The state is sync for $\gamma<\mu$ and the strongly-contrarian unsteady state for $\gamma>\mu+1$. Markers are simulation data; dashed curves are the static predictions: $r=1$, $s=\mu/\gamma$ on the cluster branches; $r=0$, $s=1$ in the phase-wave window; and $r=s=1$ on sync. The lower panel shows the mean particle speed.}
\label{fig:uniform-reentrance}
\end{figure}

Figure~\ref{fig:uniform-phase-map} shows the simulated phase diagram with the analytic boundaries overlaid: the shaded cells are the state observed from random initial conditions on a $(\mu,\gamma)$ grid, while the firm analytic async, phase-wave, and sync stability boundaries and the coupling-split existence boundaries are drawn as curves. The unsteady region is read off from the simulations (persistent drift), not assigned by elimination. Figure~\ref{fig:uniform-order-mu} shows a one-parameter section through the same diagram at $\gamma=0.5$. Figure~\ref{fig:uniform-reentrance} shows a complementary section at fixed $\mu=0.1$, sweeping $\gamma$. The coupling-split construction exists throughout $0<\mu<\gamma<\mu+1$ (here $0.1<\gamma<1.1$). The branch itself is never interrupted. However, the observed attractor from random initial conditions switches: the cluster is reached for $\gamma<\gamma_{\rm PW,lo}$ and again for $\gamma>\gamma_{\rm PW,hi}$, while the phase wave captures random trajectories in the narrow window $\gamma\in(\gamma_{\rm PW,lo},\gamma_{\rm PW,hi})$. This is a basin-level reentrance -- the cluster branch coexists with the phase wave throughout, but the phase wave is the stronger attractor in the middle window.

\begin{figure}[tb]
\centering
\includegraphics[width=\columnwidth]{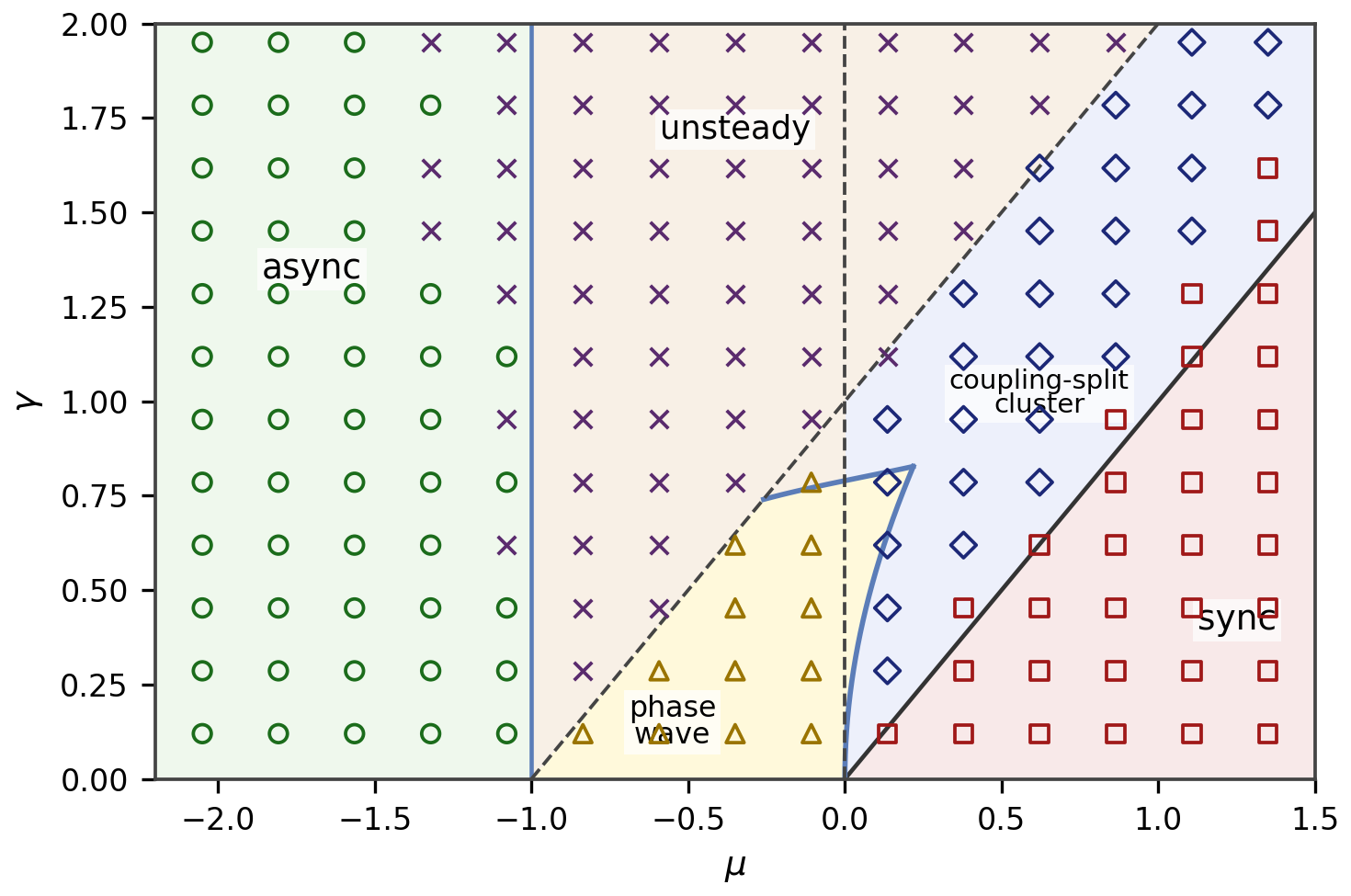}
\caption{Phase diagram in the $(\mu,\gamma)$ plane. \emph{Shaded cells} show the state observed in simulations from random initial conditions on a $40\times30$ grid ($N=112$, $dt=0.06$, $T=90$), classified as async, phase wave, coupling-split cluster, unsteady, or sync. \emph{Solid curves} are the firm analytic linear-stability boundaries: $\mu=-1$ (async), $\mu=\gamma$ (sync), and the phase-wave real-eigenvalue and Hopf curves [Eqs.~\eqref{eq:uniform-real-boundary},~\eqref{eq:uniform-osc-boundary}]. \emph{Dashed curve} is the coupling-split existence boundary $\mu=\gamma-1$ (positive-$a$ support). The analytic boundaries track the simulated state transitions; in particular the unsteady region is identified from the simulations (persistent drift), not by elimination. Isolated cells near the async edge are finite-time transients.}
\label{fig:uniform-phase-map}
\end{figure}

\section{Analysis}
\label{sec:analysis}

The stability and existence conditions derived below are summarized in Table~\ref{tab:uniform-summary}.

\begin{table*}[tb]
\caption{Summary of state conditions for $K_i'\sim U[\mu-\gamma,\mu+\gamma]$ and $\gamma\geq0$. The coupling-split row is an exact existence construction with single-particle transverse stability (\S\ref{ssec:coupling-split}); its status as the attractor reached from random initial conditions is established numerically (Figs.~\ref{fig:uniform-threshold-branch}--\ref{fig:uniform-basin}), not by a full collective stability proof.}
\label{tab:uniform-summary}
\begin{ruledtabular}
\begin{tabular}{llll}
State & Condition & Claim type & Controlling feature\\
\hline
Async & $\mu<-1$ & linear & mean $\langle a\rangle$\\
Sync & $\mu>\gamma$ & linear & lower edge $K'_{\min}>0$\\
Phase wave & $\mu-\gamma>-1$; Eq.~\eqref{eq:uniform-pw-char} stable & linear & transform of $h(K')$\\
Coupling-split cluster & $0<\mu<\gamma<\mu+1$ & exact/transverse & conformist excess\\
Strongly-contrarian unsteady & $\mu>-1,\ \gamma>\mu+1$ & empirical & no analyzed static branch\\
\end{tabular}
\end{ruledtabular}
\end{table*}

\subsection{Async}
We linearize the density $\rho$ around the uniform state $\rho_0 = (2\pi)^{-2}$, using the first-Fourier-mode projection of Strogatz and Mirollo \cite{strogatz1991stability} as adapted to the swarmalator transport equation \cite{o2022collective,o2025stability}. The full spectrum consists of a $K'$-continuum on the imaginary axis plus a single collective order-parameter eigenvalue. The first-Fourier-mode projection of the linearized transport equation gives the collective eigenvalue $\lambda = \langle a\rangle/2 = (1+\mu)/4$, so async is stable when
\begin{align}
\boxed{\mu<-1.}
\end{align}
The width $\gamma$ does not enter the order-parameter eigenvalue because the $b_i\,s\sin(\psi-\eta_i)$ term in $\dot\xi_i$ is $\xi$-independent at the uniform base $\rho_0$, so it contributes nothing to the $e^{-i\xi}$-mode of $\partial_\xi(\rho_0\dot\xi)$; only the mean of $a$ survives. The threshold $\mu=-1$ is the point at which the mean same-coordinate response $\langle a\rangle=(1+\mu)/2$ changes sign. This is distinct from the conformist--contrarian sign threshold $\langle K_i'\rangle=\mu=0$.
\subsection{Sync}
The synchronized state has $\xi_i=\xi_0$, $\eta_i=\eta_0$, hence $r=s=1$. Setting $\xi_0=\eta_0=0$ and linearizing gives
\begin{align}
\dot \xi_i &= a_i(\bar \xi-\xi_i)+b_i(\bar \eta-\eta_i),\\
\dot \eta_i &= b_i(\bar \xi-\xi_i)+a_i(\bar \eta-\eta_i).
\end{align}
The per-particle Jacobian has eigenvalues $-(a_i+b_i)=-1$ (symmetric mode) and $-(a_i-b_i)=-K_i'$ (antisymmetric mode). The symmetric mode is always damped; the antisymmetric mode is damped only when $K_i'>0$, and grows for $K_i'<0$. Thus sync is stable only if every particle is conformist, $K_i'>0$ for all $i$. For the uniform distribution this gives
\begin{align}
\boxed{\mu>\gamma.}
\end{align}
The sync threshold is set by the most contrarian particle, $K'_{\min}=\mu-\gamma$.

\subsection{Phase Wave}
\label{ssec:phase-wave}
The phase wave can be written, after a rotation of $\eta$, as
\begin{align}
\rho_0(\xi,\eta,K')=\frac{1}{2\pi}\delta(\eta),
\end{align}
where $\xi$ is uniform and $\eta$ is locked. The zeroth $\eta$ sector contains internal modes with decay rates $a_i = (1+K_i')/2$, so the first requirement is that every particle have $a_i>0$, i.e.,
\begin{align}
\mu-\gamma>-1
\label{eq:pw-positive-support}
\end{align}
(no particle so strongly contrarian that the locked $\eta$ clump destabilizes); we refer to this as the \emph{positive-$a$ support} condition to avoid confusion with positivity of $K'$.

The only non-neutral collective instability occurs in the first Fourier sector. Let $U(K',t)$ be the first Fourier moment $U(K',t)=\int e^{i\xi}\,\delta\rho_\xi(\xi,K',t)\,d\xi$ of the perturbed $\xi$ density at fixed $K'$, and let $V(K',t)=(2\pi)^{-1}\int e^{i\xi}\,y(\xi,K',t)\,d\xi$ be the first $e^{i\xi}$-moment of the small $\eta$-displacement of the locked clump; define $\bar U(t)=\int U(K',t)h(K')\,dK'$. Projecting the linearized transport equation onto these amplitudes (the factors of $i$ arise from $\partial_\xi e^{-i\xi}=-i e^{-i\xi}$) gives
\begin{align}
\dot U(K',t)&=\frac{a}{2}\bar U(t)-i\, b\, V(K',t),\\
\dot V(K',t)&=-\frac{i\, b}{2}\bar U(t)-a\, V(K',t).
\label{eq:pw-uv}
\end{align}
The same-coordinate response is $a$ and the cross-channel response is $b$. For collective modes $U,V,\bar U \propto e^{\lambda t}$, the second equation gives $V_{K'}=-i b\,\bar U/(2(\lambda+a))$, and substituting into the first and averaging over $K'$ yields the dispersion relation
\begin{align}
2\lambda
=\langle a\rangle - \int \frac{b^2(K')\,h(K')}{\lambda+a(K')}\,dK'.
\label{eq:general-pw-char}
\end{align}
The phase-wave boundary is controlled by the same-coordinate-weighted Cauchy transform of the coupling density, with both $a$ and $b$ heterogeneous. For the uniform density we change variables $u = a = (1+K')/2$ (so $b=1-u$ and the measure becomes $du/\gamma$ on $[a_-, a_+]$ with $a_\pm = (1+\mu\pm\gamma)/2$). The integrand admits the polynomial division
\begin{align}
\frac{(1-u)^2}{\lambda+u} = (u-2-\lambda) + \frac{(1+\lambda)^2}{\lambda+u},
\end{align}
which integrates in closed form to
\begin{align}
\int_{a_-}^{a_+}\!\frac{(1-u)^2}{\lambda+u}du
&= \tfrac{(1+\mu)\gamma}{2} - (2+\lambda)\gamma \notag\\
&\quad + (1+\lambda)^2\log\!\frac{\lambda+a_+}{\lambda+a_-}.
\end{align}
Dividing by $\gamma$ and substituting into \eqref{eq:general-pw-char} (with $\langle a\rangle=(1+\mu)/2$), the $(1+\mu)/2$ pieces cancel and we obtain
\begin{align}
\boxed{
\frac{(1+\lambda)^2}{\gamma}\log\!\left(\frac{1+\mu+\gamma+2\lambda}{1+\mu-\gamma+2\lambda}\right) = 2-\lambda.
}
\label{eq:uniform-pw-char}
\end{align}
In the limit $\gamma\to0$ (identical phase coupling $K'=\mu$), Eq.~\eqref{eq:uniform-pw-char} reduces to $\lambda(1+\mu) + 4\lambda^2 = 2\mu$, the identical-coupling phase-wave characteristic equation of Refs.~\cite{o2022collective,o2025stability}.

The real-eigenvalue boundary follows by setting $\lambda=0$:
\begin{align}
\boxed{
\log\!\left(\frac{1+\mu+\gamma}{1+\mu-\gamma}\right) = 2\gamma.
}
\label{eq:uniform-real-boundary}
\end{align}
This curve is the upper boundary of the phase-wave window for narrow enough distributions. There is also a finite-frequency boundary for broad enough support. Set $\lambda=i\omega$, $\omega>0$, and decompose, using the principal branch (continuous from positive real $\lambda$ in the positive-$a$ support region),
\begin{align}
\log\!\left(\frac{1+\mu+\gamma+2i\omega}{1+\mu-\gamma+2i\omega}\right)
=A(\omega)+iB(\omega),
\end{align}
where
\begin{align}
A(\omega)&=\frac{1}{2}\log\!\left[\frac{(1+\mu+\gamma)^2+4\omega^2}{(1+\mu-\gamma)^2+4\omega^2}\right],\\
B(\omega)&=\tan^{-1}\!\left(\frac{2\omega}{1+\mu+\gamma}\right) - \tan^{-1}\!\left(\frac{2\omega}{1+\mu-\gamma}\right).
\end{align}
Substituting $(1+i\omega)^2 = (1-\omega^2)+2i\omega$ into \eqref{eq:uniform-pw-char} and equating real and imaginary parts gives the Hopf boundary
\begin{align}
\boxed{
(1-\omega^2)A - 2\omega B = 2\gamma,\qquad (1-\omega^2)B + 2\omega A = -\gamma\omega.
}
\label{eq:uniform-osc-boundary}
\end{align}
Equations~\eqref{eq:uniform-real-boundary} and \eqref{eq:uniform-osc-boundary}, together with the positive-support condition \eqref{eq:pw-positive-support}, define the phase-wave region. For $0\le\gamma<\gamma_*$, the rightmost root is complex and stable inside the window, and stability is lost only through the real boundary. For $\gamma_*<\gamma<\gamma_{\rm end}$, the complex root crosses zero before the real eigenvalue does, producing a Hopf lower boundary; the window is bounded by Hopf (below) and the real boundary (above). For $\gamma>\gamma_{\rm end}$, the phase-wave window is absent.

The constants $\gamma_*$ and $\gamma_{\rm end}$ are obtained numerically from \eqref{eq:uniform-osc-boundary}; we solve to a residual of $10^{-10}$ in both real and imaginary parts and find
\begin{align}
\gamma_* &\approx 0.74002,\\
(\mu_{\rm end},\gamma_{\rm end}) &\approx (0.21836,\ 0.82728).
\end{align}
For $\gamma=0$ the full condition reduces to $-1<\mu<0$.

\subsubsection*{A Bogdanov--Takens point and the breathing coupling-split cluster}

\begin{figure}[tb]
\centering
\includegraphics[width=\columnwidth]{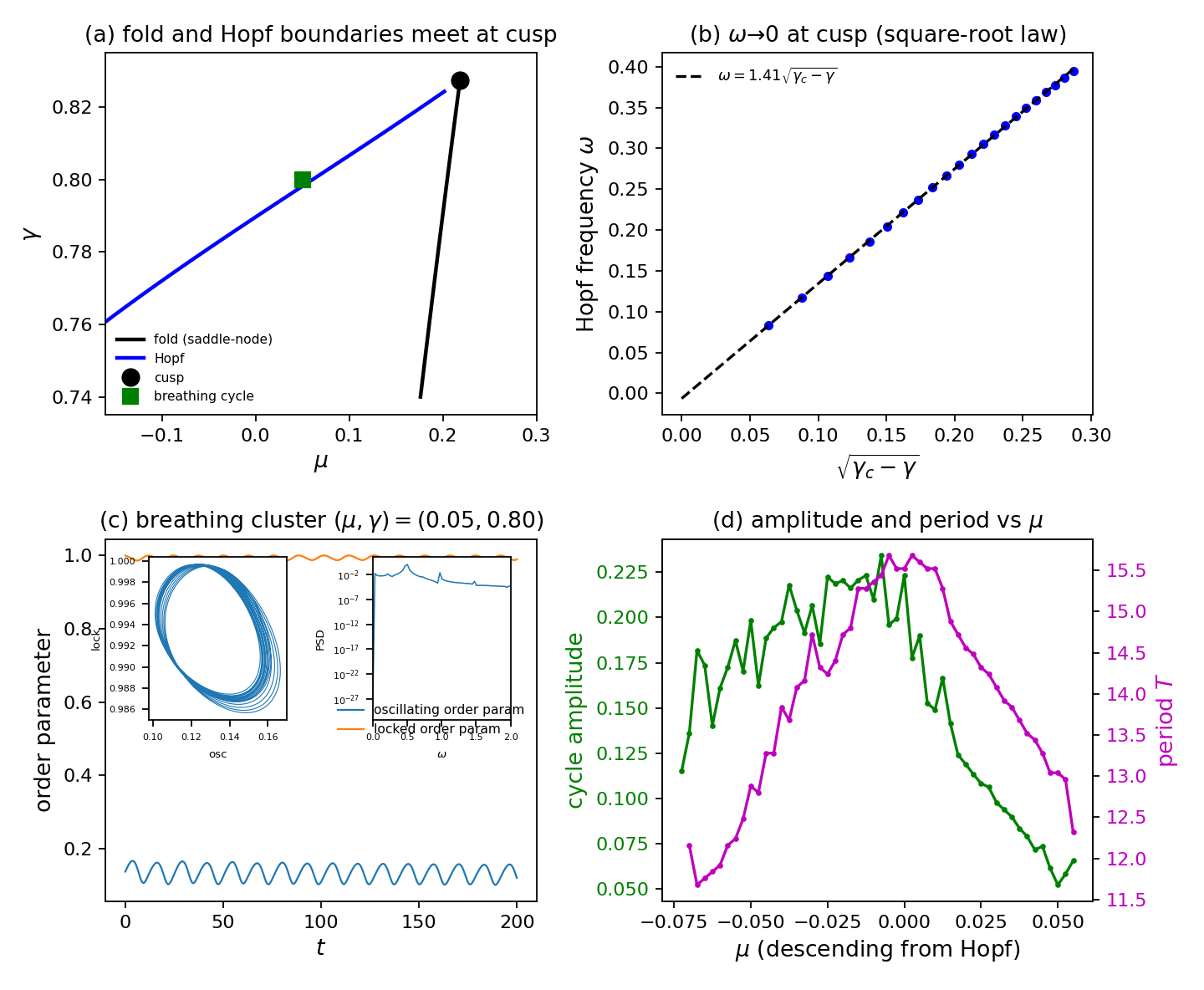}
\caption{The phase-wave cusp and its breathing limit cycle. (a)~The analytic real-eigenvalue and Hopf boundaries meet at the cusp (BT point) in the $(\mu,\gamma)$ plane; the square marks the breathing-cluster parameter point of panel~(c). (b)~The Hopf frequency vanishes as $\omega\propto\sqrt{\gamma_c-\gamma}$ at the cusp. (c)~Breathing coupling-split cluster at $(\mu,\gamma)=(0.05,0.80)$: one order parameter stays locked while the other pulses periodically; insets show the phase portrait and the single-peaked power spectrum. (d)~The cycle period grows sharply as $\mu$ decreases toward the lower edge of the limit-cycle region at $\gamma=0.80$, consistent with a global (homoclinic) bifurcation.}
\label{fig:bt}
\end{figure}

The endpoint where the Hopf and real-eigenvalue boundaries meet carries the spectral signature of a Bogdanov--Takens (BT) point in the phase-wave \emph{order-parameter dispersion}. Writing $D(\lambda;\mu,\gamma)=(1+\lambda)^2\log[(1+\mu+\gamma+2\lambda)/(1+\mu-\gamma+2\lambda)] - \gamma(2-\lambda)$, the double-zero conditions $D(0)=\partial_\lambda D(0)=0$ hold at
\begin{align}
(\mu_{\rm BT},\gamma_{\rm BT}) &\approx (0.21836,\, 0.82728),\notag\\
\partial_\lambda^2 D(0) &\approx 11.96\neq0,
\end{align}
so the marginal root has algebraic multiplicity exactly two (verified to a residual of $10^{-13}$); the locked-clump continuous spectrum lies at $\mathrm{Re}\,\lambda\in[-(1{+}\mu{+}\gamma)/2,-(1{+}\mu{-}\gamma)/2]\approx[-1.02,-0.20]$, bounded away from the axis. Consistently, the Hopf frequency vanishes at the cusp: tracking the rightmost root of $D$ along the Hopf boundary gives $\omega^2\simeq 1.99\,(\gamma_{\rm BT}-\gamma)$, i.e.\ $\omega\to0$ as $\sqrt{\gamma_{\rm BT}-\gamma}$ [Fig.~\ref{fig:bt}(b)], the square-root law expected at a BT point \cite{bogdanov1975versal,takens1974forced,guckenheimer1983nonlinear,kuznetsov2004elements}. This double zero is a property of the explicit dispersion $D$; we make no claim about a normal form for the full continuum transport problem, whose linearization also carries a neutral continuous spectrum on the imaginary axis.

Near the cusp simulations from random initial conditions show a small-amplitude limit cycle, a \emph{breathing coupling-split cluster}. At $(\mu,\gamma)=(0.05,0.80)$ one split coordinate stays locked ($s\approx1$) while the smaller order parameter $r$ pulses periodically with period $T\approx13.0$ and a clean single-frequency spectrum [Fig.~\ref{fig:bt}(c)]; the order-parameter frequency $\omega_r\approx0.48$ is twice the linear eigenmode frequency $\omega\approx0.24$, as expected since $r=|\langle e^{i\xi}\rangle|$ is a modulus. The configuration is a coupling-split cluster whose $\xi$-clumps breathe rather than hold the static value $r=|2P(K'>0)-1|$; it is shown in panel~8 of Fig.~\ref{fig:uniform-gallery}. The cycle grows continuously from the Hopf boundary ($\mu_{\rm Hopf}\approx0.06$ at $\gamma=0.80$), and as $\mu$ decreases its period grows sharply [Fig.~\ref{fig:bt}(d)], consistent with termination at a global (homoclinic) bifurcation, which we do not attempt to locate precisely. The phase wave initialized on the locked-$\eta$ branch stays there, while random initial conditions are captured by the cycle, consistent with the multistability of \S\ref{sec:numerics-states}. A boundary-value continuation study (e.g.\ \textsc{auto}/\textsc{MatCont}) of the limit-cycle and homoclinic branches is left to future work.

\subsection{Coupling-Split Clusters}
\label{ssec:coupling-split}
\textit{Existence.} The static coupling-split clusters are the new states exposed by the continuous phase-coupling distribution. Define the instantaneous force components
\begin{align}
F_{\xi i}=r\sin(\phi-\xi_i),\qquad
F_{\eta i}=s\sin(\psi-\eta_i).
\end{align}
At equilibrium,
\begin{align}
\begin{pmatrix}
a_i & b_i\\
b_i & a_i
\end{pmatrix}
\begin{pmatrix}
F_{\xi i}\\
F_{\eta i}
\end{pmatrix}
=0.
\end{align}
The determinant of this matrix is $a_i^2-b_i^2 = K_i'$, so for all particles with $K_i'\neq 0$ the matrix is invertible and
\begin{align}
F_{\xi i}=F_{\eta i}=0.
\end{align}
Thus each particle must sit either with or opposite each mean field:
\begin{align}
\xi_i\in\{\phi,\phi+\pi\},\qquad
\eta_i\in\{\psi,\psi+\pi\}.
\end{align}

This immediately gives a general coupling-split construction for any normalized density $h(K')$ whose support straddles $K'=0$ (i.e., contains both conformist and contrarian particles). Let
\begin{align}
p_+=\int_0^\infty h(K')\,dK',\qquad
p_-=\int_{-\infty}^0 h(K')\,dK'.
\end{align}
If $p_+>p_-$, there is a $\xi$-dominant branch in which the conformist ($K'>0$) particles align in $\eta$ and the contrarian ($K'<0$) particles anti-align in $\eta$:
\begin{align}
r=1,\qquad s=p_+-p_-=2p_+-1.
\label{eq:general-threshold-order}
\end{align}
The coordinate-swapped branch has $r=2p_+-1$ and $s=1$. The split-coordinate order parameter is not controlled by a moment of $h(K')$; it is controlled by the conformist excess -- the mass imbalance across the sign threshold $K'=0$.

Assume the support straddles $K'=0$ and the most contrarian particle still has $a>0$:
\begin{align}
\mu-\gamma<0<\mu+\gamma, \qquad \mu-\gamma > -1.
\label{eq:threshold-support}
\end{align}
For the $\xi$-dominant coupling-split cluster, all particles align in $\xi$, while their $\eta$ coordinate is selected by the sign of $K_i'$:
\begin{align}
\xi_i&=\phi,\\
\eta_i&=
\begin{cases}
\psi, & K_i'>0,\\
\psi+\pi, & K_i'<0.
\end{cases}
\label{eq:threshold-eta}
\end{align}
This gives
\begin{align}
r=1,\qquad
s=P(K'>0)-P(K'<0).
\end{align}
For the uniform distribution,
\begin{align}
P(K'>0)=\frac{\mu+\gamma}{2\gamma},
\qquad
P(K'<0)=\frac{\gamma-\mu}{2\gamma}.
\end{align}
Self-consistency of the mean-field phase in \eqref{eq:threshold-eta} requires the aligned (conformist) clump to have larger mass, so $\mu>0$. In that case
\begin{align}
\boxed{
r=1,\qquad s=\frac{\mu}{\gamma}.
}
\label{eq:threshold-order}
\end{align}
The coordinate-swapped branch has
\begin{align}
\boxed{
r=\frac{\mu}{\gamma},\qquad s=1.
}
\end{align}
In the original position--phase variables,
\begin{align}
x_i=\frac{\xi_i+\eta_i}{2},\qquad
\theta_i=\frac{\xi_i-\eta_i}{2}.
\end{align}
Thus the $\xi$-dominant coupling-split cluster is a two-clump state in which positively and negatively coupled particles share the same $\xi=x+\theta$ coordinate but occupy opposite $\eta=x-\theta$ clumps. The coupling-split cluster exists as a static solution throughout the region
\begin{align}
\boxed{0<\mu<\gamma<\mu+1.}
\label{eq:threshold-region}
\end{align}
The three inequalities encode: positive-coupling majority ($\mu>0$), support straddles $K'=0$ ($\gamma>\mu$), and no strongly negatively coupled particle ($\gamma<\mu+1$). The coordinate-swapped branch has the same condition.

\emph{Transverse stability.} Linearizing the single-particle dynamics \eqref{eq:uniform-xi}--\eqref{eq:uniform-eta} about each clump position at frozen mean fields $(r,\phi,s,\psi)$ gives the Jacobian
\begin{align}
J_i=-\begin{pmatrix} a_i r\sigma_{\xi} & b_i s\sigma_{\eta}\\[2pt] b_i r\sigma_{\xi} & a_i s\sigma_{\eta}\end{pmatrix},
\end{align}
where $\sigma_{\xi},\sigma_{\eta}=\pm1$ record whether the particle aligns ($+$) or anti-aligns ($-$) with each mean field. Its determinant and trace are
\begin{align}
\det J_i = \sigma_{\xi}\sigma_{\eta}\,rs\,K_i',\qquad
\mathrm{tr}\,J_i = -a_i\,(r\sigma_{\xi}+s\sigma_{\eta}).
\end{align}
For the assignment in \eqref{eq:threshold-eta} -- conformists aligned in both coordinates and contrarians anti-aligned in $\eta$ -- one has $\sigma_{\xi}\sigma_{\eta}=\mathrm{sign}(K_i')$, so $\det J_i = rs\,|K_i'|>0$, while $\mathrm{tr}\,J_i=-a_i(r+s)<0$ for conformists and $-a_i(r-s)<0$ for contrarians (using $r=1$, $0<s<1$). Both eigenvalues thus have negative real part for every particle precisely when $a_i=(1+K_i')/2>0$ for all $i$, i.e.\ when $\gamma<\mu+1$ -- the same bound as existence. This is a single-particle (frozen-mean-field) result; the collective stability of the branch is left open and supported only numerically below.

\begin{figure}[!htbp]
\centering
\includegraphics[width=\columnwidth]{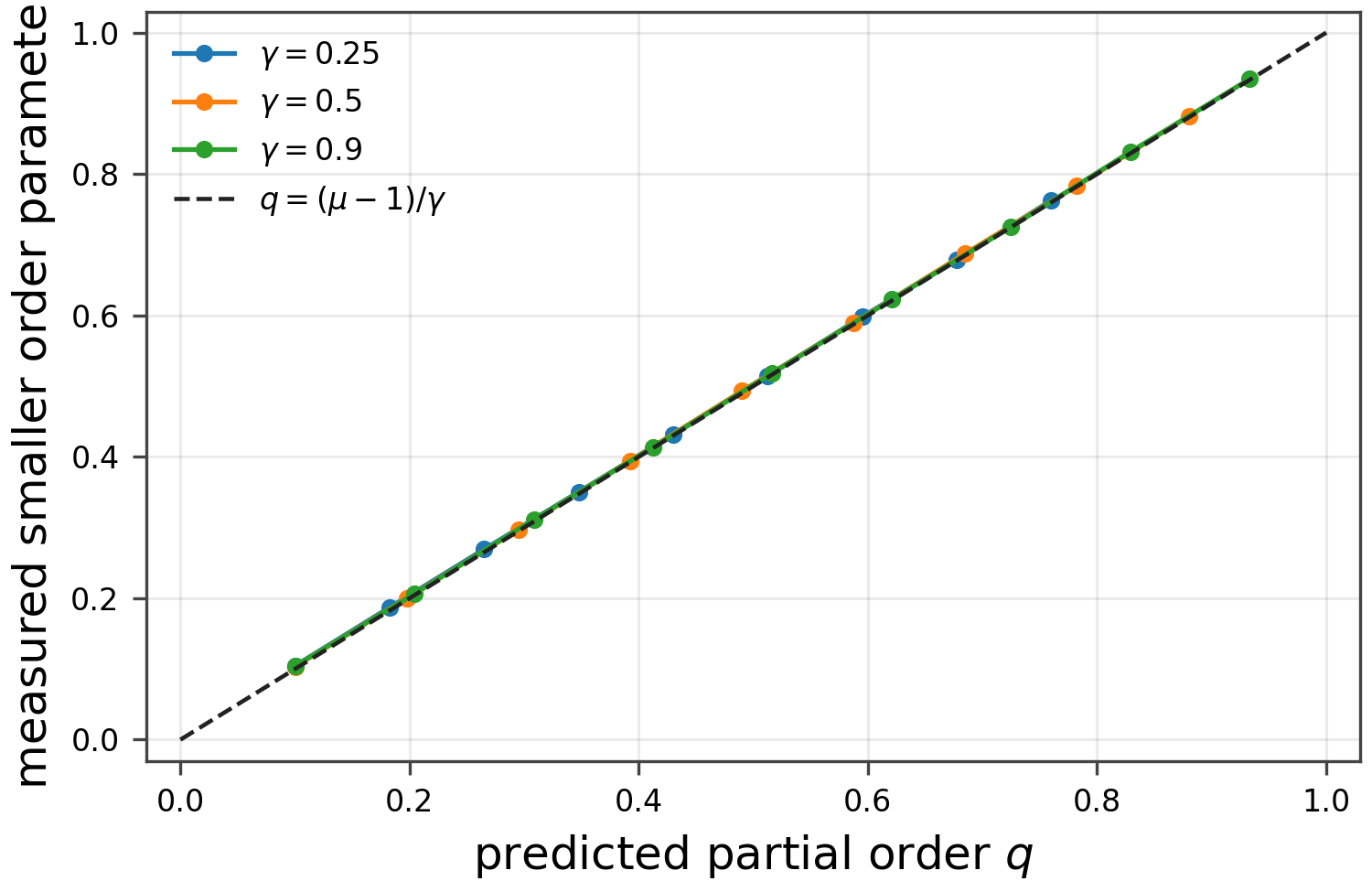}
\caption{Numerical check of the coupling-split order parameter. The measured smaller order parameter agrees with the formula $s=\mu/\gamma$ for several values of $\gamma$. Each point is a finite-$N$ simulation ($N=160$) initialized near the coupling-split branch.}
\label{fig:uniform-threshold-branch}
\end{figure}

\begin{figure}[!htbp]
\centering
\includegraphics[width=\columnwidth]{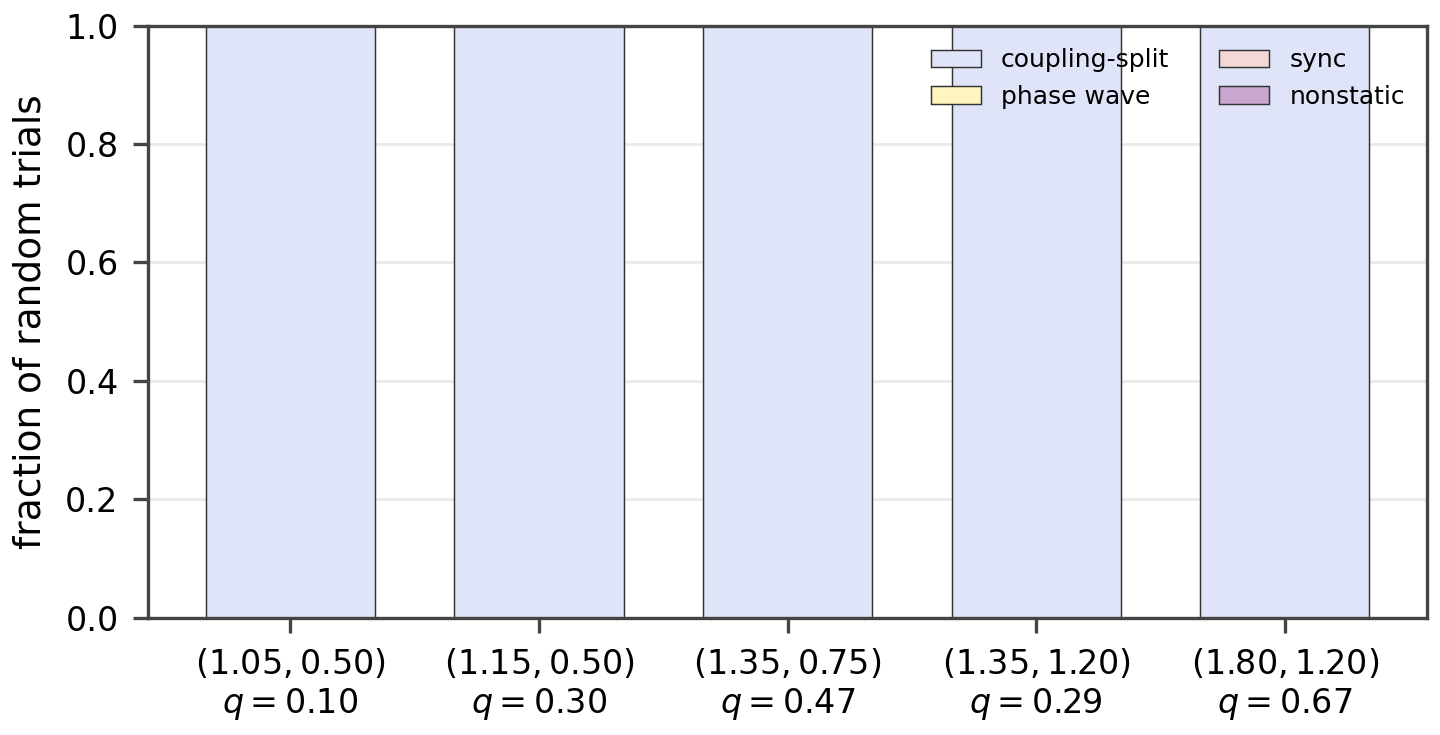}
\caption{Random-initial-condition basin census inside the coupling-split region. Each bar summarizes 64 independent random initial conditions ($N=160$, $T=120$). The displayed $s$ is the theoretical smaller order parameter $s=\mu/\gamma$.}
\label{fig:uniform-basin}
\end{figure}

Figure~\ref{fig:uniform-threshold-branch} confirms the prediction $s=\mu/\gamma$. For $(\mu,\gamma)=(0.20,0.50)$, the measured $s$ was $0.401$--$0.403$ across $N=80,160,320$, compared with the prediction $s=0.400$. For $(\mu,\gamma)=(0.35,0.70)$, values were $0.500$--$0.503$, compared with $s=0.500$. The state is reached from random initial conditions throughout the region (Fig.~\ref{fig:uniform-basin}).

The equal-weight binary split state is the endpoint $s=0$ of this picture. For a binary distribution with one mass of contrarians and one mass of conformists in equal weight, the construction gives no majority clump and hence zero partial order in the split coordinate. For the continuous uniform distribution with $\mu>0$, the conformist excess across $K'=0$ makes the partial order nonzero and turns the coupling-split cluster into an observable partially ordered state in simulations.

\subsection{Strongly-Contrarian Unsteady Dynamics}
\label{ssec:ki-born}
\begin{figure}[tb]
\centering
\includegraphics[width=\columnwidth]{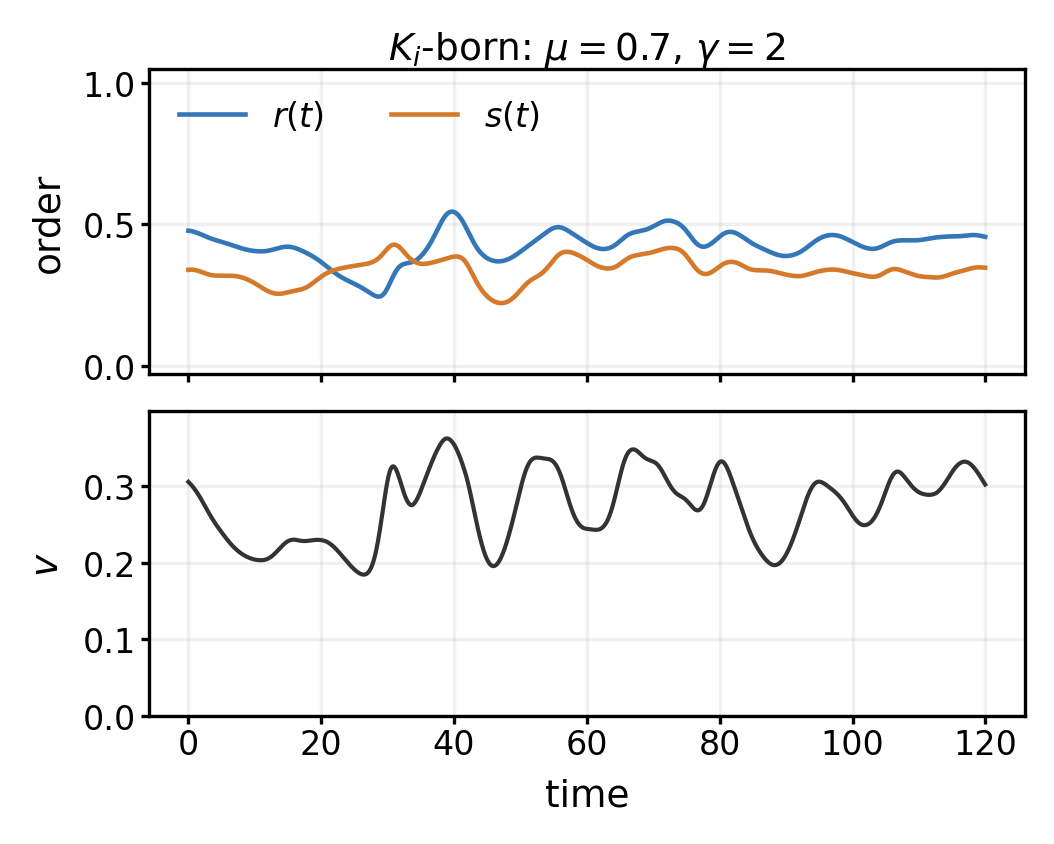}
\caption{Strongly-contrarian unsteady dynamics at $\mu=-0.3$, $\gamma=1.5$. Top: time series of $r(t)$ and $s(t)$ over a 200-unit window. Both order parameters wander over a wide range. Bottom: power spectral density $P_r(\omega)$ and $P_s(\omega)$ computed by Welch's method over $T=1500$ time units after a long transient ($N=512$, $dt=0.03$). The spectrum is broadband, distinguishing this dynamics from the small-amplitude periodic orbit of \S\ref{ssec:phase-wave}, whose PSD would be dominated by a single line.}
\label{fig:ki-nonstatic}
\end{figure}

The model supports two qualitatively distinct forms of nonstatic behavior. The first is the small-amplitude periodic orbit of \S\ref{ssec:phase-wave}: $r$ oscillates with period $T\approx 13$ in the limit-cycle wedge born at the BT cusp $(\mu_{\rm BT},\gamma_{\rm BT})\approx(0.218,0.827)$. The second, described below, is qualitatively different: it has a broadband power spectrum (Fig.~\ref{fig:ki-nonstatic}, lower panel), large amplitude, and a wide wedge of occupancy in $(\mu,\gamma)$.

Simulations in the strongly-contrarian regime reveal this second unsteady dynamics. It is produced by the phase-coupling distribution itself, without delay or external forcing. Empirically it occupies the wedge
\begin{align}
\boxed{\mu>-1,\qquad \gamma>\mu+1,}
\label{eq:ki-born-region}
\end{align}
where the support $[\mu-\gamma,\mu+\gamma]$ contains at least one strongly contrarian particle ($K_i'<-1$) while the mean same-coordinate response stays positive, $\langle a\rangle=(1+\mu)/2>0$ (i.e.\ $\mu>-1$, so that async is unstable). Note the mean coupling $\langle K_i'\rangle=\mu$ may itself be negative here. In this region none of the four static branches analyzed above is admissible: async is linearly unstable because $\mu>-1$; the phase wave is not admissible because the positive-support condition $\mu-\gamma>-1$ fails; sync is unreachable because $\mu<\gamma$; and the coupling-split construction of Eq.~\eqref{eq:threshold-region} is unavailable because its $\gamma<\mu+1$ bound is violated. We do not claim that no static branch exists in the wedge. The pointwise equilibrium condition $F_{\xi i}=F_{\eta i}=0$ still admits a combinatorial family of four-corner sign assignments, and the transverse Jacobian determinant computed in \S\ref{ssec:coupling-split} is $\sigma_\xi\sigma_\eta\,rs\,K_i'$ at each particle, so additional sub-threshold static branches involving particles with $K_i'<-1$ could in principle exist. A complete enumeration of those branches is beyond the scope of this paper; what we establish here is that among the analyzed branches none is attracting in the wedge, and simulations from random initial conditions throughout the wedge converge to unsteady dynamics rather than to any of them.

Figure~\ref{fig:ki-nonstatic} shows a representative time series at $\mu=-0.3$, $\gamma=1.5$. Both $r(t)$ and $s(t)$ oscillate persistently with a nontrivial waveform, and the mean particle speed $v$ stays close to $0.23$ throughout. The state is reached from random initial conditions in our $(\mu,\gamma)$ scan (Fig.~\ref{fig:uniform-phase-map}), so it is not a metastable transient on the scan timescale. The broadband spectrum and a positive largest Lyapunov exponent suggest irregular dynamics: a Benettin calculation (two nearby trajectories, separation renormalized every $0.2$ time units after an $800$-unit transient, $N=400$, $T=4000$) gives $\lambda_{\max}\approx0.11>0$, distinguishing this dynamics from the single-frequency breathing cycle of \S\ref{ssec:phase-wave}, for which the same estimate is consistent with zero. The leading exponent is unaffected by the two neutral rotational modes.

The empirical boundaries of this region are themselves transitions of the analyzed static branches. The lower boundary $\mu=-1$ is the async destabilization, on which the leading eigenvalue $\lambda=(1+\mu)/4$ crosses zero through a real eigenvalue rather than a Hopf. Thus the unsteady dynamics is not produced by a Hopf of async; it is observed after the analyzed static attractors are absent above $\mu=-1$ in the strongly-contrarian wedge. The boundary $\gamma=\mu+1$ is the positive-support line of the phase wave (equivalently, the upper edge of the coupling-split region); crossing it from the wedge into $\gamma<\mu+1$ restores at least one of the analyzed static attractors (phase wave or coupling-split cluster, depending on $\mu$) and the oscillations disappear. A finer characterization of the spectrum and amplitude of this dynamics, and a systematic enumeration of strongly-contrarian static branches, are left for future work.

\subsection{Robustness to Compact-Support Densities}
The coupling-split construction does not rely on the uniform density. Let $g(K')$ be any compactly supported density with support straddling $K'=0$ and contained within $(-1,\infty)$. Then the same pointwise equilibrium condition gives
\begin{align}
r=1,\qquad s=q_g=\left|2P_g(K'>0)-1\right|,
\label{eq:robust-q}
\end{align}
provided the larger subpopulation is used as the aligned clump. The uniform distribution is the special case $q_g=\mu/\gamma$.

\begin{figure}[tb]
\centering
\includegraphics[width=\columnwidth]{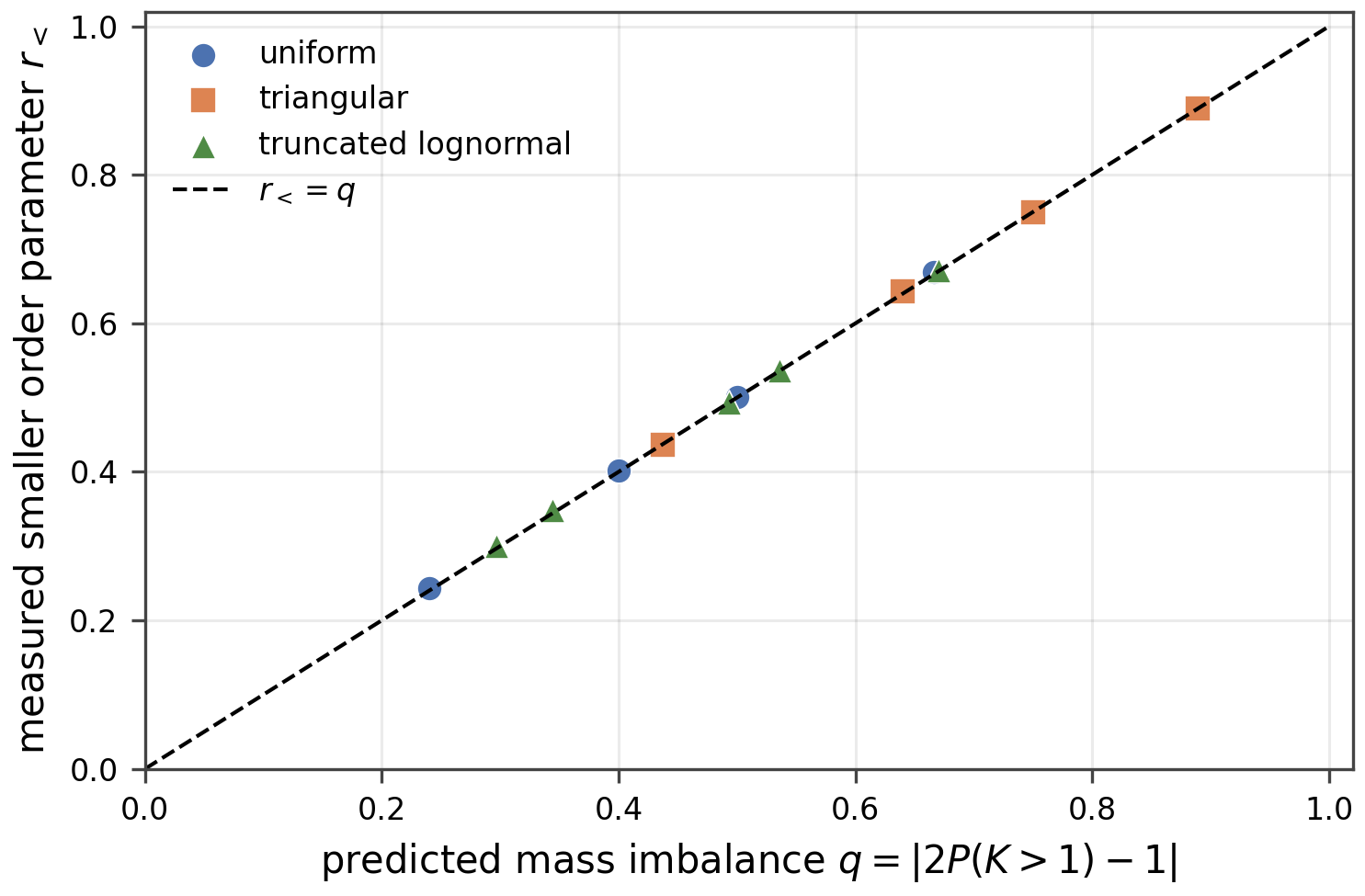}
\caption{Robustness of the coupling-split order parameter for compactly supported phase-coupling densities. Each point is a finite-$N$ simulation initialized near the coupling-split branch. The measured smaller order parameter $s$ collapses onto the conformist-excess prediction $q_g=|2P_g(K'>0)-1|$ for uniform and triangular distributions.}
\label{fig:compact-robustness}
\end{figure}

As a first non-uniform example, take the symmetric triangular density centered at $\mu$ with half-width $\gamma$. For $\delta=\mu/\gamma\in(0,1)$ and $\mu>0$, Eq.~\eqref{eq:robust-q} gives
\begin{align}
q_\triangle=2\delta-\delta^2.
\end{align}
Figure~\ref{fig:compact-robustness} shows that finite-$N$ simulations initialized near the coupling-split branch collapse onto Eq.~\eqref{eq:robust-q} for uniform and triangular couplings (the two symmetric compactly-supported families with closed-form $q_g$). Asymmetric compactly-supported families are expected to give the same collapse, since Eq.~\eqref{eq:robust-q} depends only on the conformist mass and not on the shape of $h(K')$; we have not exhaustively tested asymmetric densities here because samples placing mass close to $K'=0$ produce particles whose transverse relaxation timescale $\propto 1/|K_i'|$ diverges, which spoils the cluster construction on the simulation timescale rather than the formula itself.

This robustness test concerns the coupling-split branch, not the full phase diagram. The async state depends on $\langle K'\rangle$, sync depends on the lower edge of the support, and the phase-wave boundary depends on the full transform $\int g(K')(2\lambda+1+K')^{-1}\,dK'$. The compact-support restriction is important: if the support extends arbitrarily close to $K'=0$ or down to $K'=-1$, the spectral gap associated with the split construction can collapse.

\section{Discussion}
We studied the 1D swarmalator model with phase coupling $K_i'$ drawn from a uniform distribution. The main result is simple: each familiar state survives, but feels a different part of the coupling spread -- async the mean response $\langle a\rangle=(1+\mu)/2$, sync the most contrarian particle $\mu-\gamma$, and the phase wave the full density through a logarithmic characteristic equation [Eq.~\eqref{eq:uniform-pw-char}]. Straddling the sign threshold $K'=0$ adds a new static state, the coupling-split cluster, in which positively and negatively coupled particles occupy antipodal clumps; its order parameter is set by the conformist excess alone, reducing for the uniform law to $s=\mu/\gamma$.

What is interesting is that this cluster is invisible in the binary conformist--contrarian problem, where the equal-weight split gives $s=0$ and the state is marginal. The continuous distribution does more than smooth the binary phase diagram: its nonzero conformist excess turns the split into an observable partially ordered family. Two further features sit at the edges of the static picture -- the phase-wave dispersion has a Bogdanov--Takens cusp with a nearby small-amplitude breathing cycle, and strongly contrarian supports ($\gamma>\mu+1$) drive persistent, irregular oscillations of the order parameters.

Future work could pin down what we left numerical: a continuum stability spectrum for the coupling-split branch, whose modes accumulate near $K'=0$; the attractors of the strongly-contrarian wedge and any static branches hidden there; and a continuum normal form for the BT cusp. Adding more realism -- broader coupling densities, distributed natural frequencies, delay, or two spatial dimensions \cite{o2024solvable} -- may also be fruitful.

\begin{acknowledgments}
The author thanks the members of the Starling Research Institute for helpful discussions.
\end{acknowledgments}

\section*{Data availability}
The code used to generate the simulations and figures in this paper is available from the author on reasonable request.

\appendix
\section{iid-disorder robustness checks}
\label{app:iid}

The main-text scan uses deterministic quantiles $K_i'=\mu-\gamma+2\gamma(i+1/2)/N$ as a quadrature rule for the continuum distribution. To verify that the observed states are not artifacts of this choice, we repeated representative simulations with iid quenched uniform couplings ($K_i'$ drawn independently for each particle). Table~\ref{tab:random-disorder-check} shows that the phase wave, coupling-split cluster, and strongly-contrarian unsteady dynamics all persist across 20 independent disorder realizations.

\begin{table*}[tb]
\caption{Representative iid-disorder checks at $N=256$. Random columns summarize 20 independent quenched draws of $K_i'$. For the coupling-split row, $q_{\rm samp}=|N^{-1}\sum_i {\rm sign}(K_i')|$ is the realized finite-sample conformist excess.}
\label{tab:random-disorder-check}
\begin{ruledtabular}
\begin{tabular}{lll}
Test point & Quantiles & iid quenched $K_i'$\\
\hline
Phase wave $(-0.30,0.20)$ & phase wave & 20/20 phase wave\\
Coupling split $(0.20,0.50)$ & $s=0.402$ & 20/20 split; $s=0.427\pm0.053$, $q_{\rm samp}=0.414\pm0.058$\\
Strong contrarian $(-0.30,1.50)$ & unsteady; $v=0.237$ & 20/20 unsteady; $v=0.233\pm0.008$\\
\end{tabular}
\end{ruledtabular}
\end{table*}

\bibliographystyle{apsrev4-1}
\bibliography{main_uniform_Ki_v3}

\end{document}